\documentclass[aps,prb,twocolumn,superscriptaddress,showkeys,showpacs,longbibliography,groupedaddress]{revtex4-1}
\usepackage[utf8]{inputenc}
\usepackage{graphicx} 
\usepackage{amssymb}
\usepackage{amsmath}
\usepackage{dcolumn} 
\DeclareMathVersion{nxbold}
\SetSymbolFont{operators}{nxbold}{OT1}{cmr} {b}{n}
\SetSymbolFont{letters}  {nxbold}{OML}{cmm} {b}{it}
\SetSymbolFont{symbols}  {nxbold}{OMS}{cmsy}{b}{n}
\makeatletter
\newcolumntype{d}[1]{D{.}{.}{#1} } 
\newcolumntype{Z}[3]{>{\mathversion{nxbold}\DC@{#1}{#2}{#3} }c<{\DC@end} } 
\makeatother

\begin{document}
\title{Finite-size scaling of Monte Carlo simulations for the fcc Ising antiferromagnet: Effects of the low-temperature phase degeneracy}
\author{Ronja St\"ubel}
\email{ronja.stuebel@itp.uni-leipzig.de}
\author{Wolfhard Janke}
\email{wolfhard.janke@itp.uni-leipzig.de}
\affiliation{Institut für Theoretische Physik, Universität Leipzig, Postfach 100\,920, 04009 Leipzig, Germany}
\date{\today}
 \begin{abstract}
The Ising antiferromagnet on a face-centered cubic (fcc) lattice with nearest-neighbor interaction only is well known to exhibit a macroscopic (exponential in the system size $L$) ground-state degeneracy. With increasing temperature, this degeneracy is expected to be lifted and the model undergoes a first-order phase transition. For a model with an exponential degeneracy in the {\em whole\/} low-temperature phase, it was recently found that the finite-size scaling behavior is governed by leading correction terms $\sim L^{-2}$ instead of $\sim L^{-3}$ as usual. To test the conjecture that such a transmuted behavior may effectively persist also for the fcc antiferromagnet up to some crossover system size, we have performed parallel multicanonical Monte Carlo simulations for lattices of linear size $L \le 18$ with periodic boundary conditions and determined various inverse pseudo phase transition temperatures, as well as the extremal values of the specific heat and the energetic Binder parameter. We indeed find that, for the simulated lattice sizes, the conjectured transmuted finite-size scaling ansatz fits the data better than the standard ansatz. On this basis, we extrapolate for the transition temperature an estimate of $T_0 = 1.735047(46)$.
\end{abstract}
\maketitle

\section{Introduction}
The ordering of the Ising antiferromagnet on a face-centered cubic (fcc) lattice presents a long-standing problem which has received extensive attention since the 1930s.~\cite{Peierls1936, Shockley1938, Luttinger1951, Danielian1964, Betts1965, VanBaal1973,  Slawny1979, Phani1979, Phani1980, Binder1980, Alexander1980,  Mackenzie1981, Lebowitz1985, Kaemmerer1996, Beath2005, Beath2006} Initially, the model was employed and examined as an approximation of ordering binary alloys. For sufficiently small exterior magnetic field $|h|$ and ferromagnetic or vanishing next-nearest neighbor interaction, the fcc Ising antiferromagnet can describe magnetic alloys with the metallurgist's ``AB'' or ``L1$_0$" structure~\cite{Lebowitz1985,Bricmont1989} which are currently widely studied since they represent promising materials for the heat-assisted magnetic recording (HAMR) technology for ultra-high density magnetic recording media.~\cite{Laughlin2005, Graf2007, Wang2011, Hono2016}

Yet another motivation for the research on the fcc Ising antiferromagnet is the interest in frustrated magnetism in general which mainly prompted the later publications on this model.~\cite{Binder1980, Alexander1980, Mackenzie1981, Kaemmerer1996, Beath2005, Beath2006} Due to the conjunction of the antiferromagnetic nearest-neighbor interaction and the geometry of the fcc lattice, it is impossible to satisfy all interaction bonds of the fcc Ising antiferromagnet simultaneously which is referred to as geometrical frustration.~\cite{Toulouse1977} Frustrated systems are subject to numerous investigations since they, in general, give rise to interesting, complex properties while being difficult to solve.~\cite{Beath2005} Besides, the behavior of frustrated systems is hoped to shed some light on spin glasses which, too, show frustration but are even more complicated since they, in addition, involve randomness.

\begin{figure}
\includegraphics{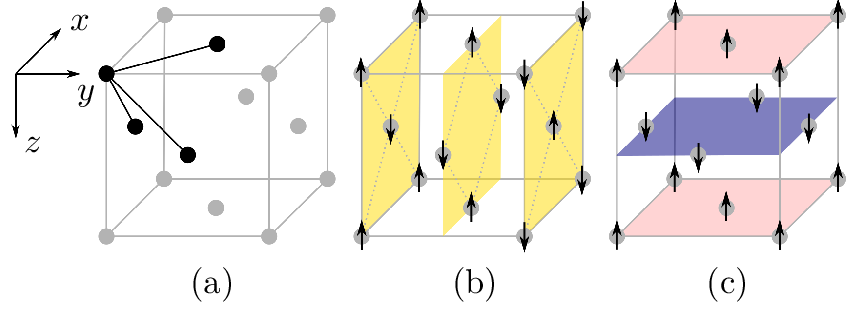}
 \caption{\label{fig:L_and_groundstate}(a)~The linear system size $L$ is defined as the number of the depicted four-spin-complexes (black) along the $x$-, $y$-, or $z$-axis. (b)~Exemplary illustration of a ground state. The shaded planes are ordered antiferromagnetically. (c)~One of the AB structure ground states.}
\end{figure}

Here, the attention is dedicated to the model with vanishing external magnetic field $h=0$ and nearest-neighbor interaction only. That is, the system given by the Hamiltonian
\begin{equation}
\label{eq:Hamiltonian}
\mathcal{H}=-J \sum\limits_{\langle i,j\rangle} s_i s_j 
\end{equation}
is regarded where the sum ranges over the nearest-neighbor Ising spins $s_i=\pm 1$ on a fcc lattice with periodic boundary conditions and $J<0$ denotes the coupling constant. Only systems with the same linear dimension $L$ in the $x$-, $y$-, and $z$-direction are considered and $L$ is defined as the number of the four-spin-complexes depicted in Fig.~\ref{fig:L_and_groundstate}(a) along the $x$-, $y$-, or $z$-direction yielding a total number of $V=4L^3$ lattice sites. This Hamiltonian \eqref{eq:Hamiltonian} is particularly worthy of consideration since its ground state features only a two-dimensional long-range order and thus is infinitely degenerate in the thermodynamic limit.~\cite{Luttinger1951} However, as soon as the temperature is lifted above zero, the infinitely large system exhibits ``order out of disorder"~\cite{Villain1980} leading to a three-dimensional long-range order.~\cite{Bricmont1989} 

The exploration of the model \eqref{eq:Hamiltonian} poses two major challenges since it is geometrically frustrated and shows a first-order phase transition. As a consequence of that, disagreeing results for the model have been obtained in the past.~\cite{Bragg1934, Bethe1935, Li1949, Kikuchi1974, Binder1980} The first high-precision estimate of the phase transition temperature could be determined by Beath and Ryan in 2006.~\cite{Beath2006} Interestingly, their simulation data suggests that the model does not comply with the standard scaling $\sim L^{-3}$ for a common inverse pseudo phase transition temperature $\beta_0(L)$ at a first-order phase transition. However, they could not substantiate this behavior theoretically. On the other hand, in 2014, Mueller et al.~\cite{Mueller2014b, Mueller2014, Mueller2014a} deduced that the non-standard scaling $\sim L^{-2}$ is expected to apply to models where the number $q$ of low-temperature ordered phases grows exponentially with the linear system size $L$. This has numerically been confirmed for the three-dimensional purely plaquette gonihedric Ising model.~\cite{Mueller2014} Since for the fcc Ising antiferromagnet \eqref{eq:Hamiltonian}, at least the ground-state degeneracy is also exponential in $L$, the question arises whether this model behaves in a similar way. According to Mueller et al.\cite{Mueller2014,Mueller2014a}, it is conceivable that a crossover from the non-standard scaling $\sim L^{-2}$ to the standard scaling $ \sim L^{-3}$ might be observed when increasing the lattice size $L$.

Here, the open question regarding the scaling of $\beta_0(L)$ is investigated by means of parallelized Monte Carlo simulations in multicanonical ensembles which provide the benefit of bypassing the typical problems of canonical simulations close to first-order phase transitions. 

The rest of the paper is organized as follows: In section~\ref{sec:Model_and_methods}, established knowledge about the model is reviewed and the applied methods are explicated. Subsequently, the numerical implementation and results are presented in section~\ref{sec:Numerical_results}. The conclusion is given in section~\ref{sec:Conclusion}.

\section{Model and methods} \label{sec:Model_and_methods}

\subsection{Model} \label{sec:Model}
In contrast to the simple cubic (sc) and the body-centered cubic (bcc) lattice, the fcc lattice is not bipartite. That means, it cannot be divided into two parts such that all nearest neighbors of one part belong to the other part and vice versa. As a consequence, it is impossible to order the fcc lattice completely antiferromagnetically. Thus, the system \eqref{eq:Hamiltonian} is geometrically frustrated resulting in a large number of ground states. The ground-state properties have already been successfully analytically investigated.~\cite{Luttinger1951, Danielian1961, Danielian1964} Each ground state consists of a stack of uncorrelated, antiferromagnetically ordered planes which are parallel to the $xy$-, $xz$-, or $yz$-plane, see Fig.~\ref{fig:L_and_groundstate}(b). There are six equivalent~\footnote{Two spin configurations are called equivalent to each other if they are related  by the symmetries of the Hamiltonian.} ground states which possess a higher symmetry than all the other ground states. They are composed of alternating layers of alike spins, see Fig.~\ref{fig:L_and_groundstate}(c). This corresponds to the ``AB'' or ``L1$_0$" structure in the alloy analog. Among all ground states, these AB structure ground states possess the highest density of low-energy excitations and thus are called dominant.~\footnote{A formal definition of a dominant ground state is given in Ref.~\onlinecite{Slawny1979}}
Providing for the high symmetry of these six AB structure ground states, the ground-state degeneracy
\begin{equation}
g_0 = 3 \times 2^{2L} - 6
\label{eq:groundstate_degeneracy}
\end{equation}
is derived since there are three possible orientations of the stack of $2L$ antiferromagnetically ordered planes and each plane can be ordered antiferromagnetically in two different ways. That means, the ground-state degeneracy, which defines the number $q$ of ordered phases at zero temperature, is exponential in $L$. However, for the infinitely large system, only the six dominant AB structure ground states survive the transit to (small) temperatures above zero which was rigorously proven by Bricmont and Slawny.~\cite{Bricmont1989} Unfortunately, the number $q$ of ordered phases for temperatures $T>0$ and finite lattice sizes $L$ is unknown. Following the considerations of K\"ammerer et al.,~\cite{Kaemmerer1996} it might be constant for sufficiently large lattice sizes while depending exponentially on $L$ for small systems.

The model~\eqref{eq:Hamiltonian} shows a first-order phase transition where the system goes from a $q$-degenerate ordered antiferromagnetic phase at low temperatures to a disordered paramagnetic phase at higher temperatures. Characteristically for a first-order phase transition, these different phases coexist at the phase transition temperature which is reflected in a double peak structured canonical energy probability distribution. 

\subsection{Methods} \label{sec:Methods}

\subsubsection{Finite-size scaling for first-order phase transitions}
Let us first recall a simple approach for describing the behavior of a first-order phase transition: the two-state ansatz.~\cite{Janke1993, Janke2003} Here, all fluctuations within the pure phases (ordered and disordered) are neglected. Thus, the time evolution of a system near the phase transition temperature shows sharp jumps between two possible values $u_\text{o} = \frac{\mathrm{d} \left( \beta f_\text{o} \right)}{\mathrm{d}\beta}$, $u_\text{d}= \frac{\mathrm{d} \left( \beta f_\text{d} \right)}{\mathrm{d}\beta}$ of the energy per lattice site $e$, where $f_\text{o}$, $f_\text{d}$ define the infinite-volume free energy densities associated with the ordered phases and the disordered phase, respectively, and $\beta=1/k_B T$ denotes the inverse temperature, which is related to the temperature $T$ via the Boltzmann constant $k_B$. Using $f_\text{o}$ and $f_\text{d}$, one can calculate the fraction $W_\text{o} \propto q e^{-\beta f_\text{o}V}$ of time spent in the ordered phases (corresponding to $u_\text{o}$) and the fraction $W_\text{d} = 1- W_\text{o} \propto e^{-\beta f_\text{d}V}$ of time spent in the single disordered phase (corresponding to $u_\text{d}$) for $L\to \infty$. This directly gives the energy moments $\langle e^n \rangle = W_\text{o} u^n_\text{o} + W_\text{d} u^n_\text{d}$ for all $n \in \mathbb{N}$. The heat capacity per lattice site (specific heat) can then be expressed as $c=\beta^2 V \left( \langle e^2 \rangle - \langle e \rangle^2  \right) = \beta^2 V W_\text{o} (1-W_\text{o})\Delta u^2$ where $\Delta u = u_\text{d}- u_\text{o}$. It takes the maximal value $c_\text{max} \approx \beta_0^2 V (\Delta \hat{u}/2)^2$ for $W_\text{o} \approx W_\text{d}$. That means, in this two-state ansatz, the location $\beta_{c_\text{max}}(L)$ of the specific heat maximum and the inverse temperature $\beta_\text{eqw}(L)$, where the ordered and disordered peak of the canonical energy probability distribution have the same weight, approximately coincide. One can calculate these inverse pseudo phase transition temperatures $\beta_{c_\text{max}}(L)$, $\beta_\text{eqw}(L)$ by taking the logarithm of the ratio $\frac{W_\text{o}}{W_\text{d}} \approx 1 \approx q e^{\beta (f_\text{d}-f_\text{o})V}$ and performing a Taylor expansion around the inverse phase transition temperature $\beta_0$. This results in the finite-size scaling formula $\beta_{c_\text{max}}(L) \approx \beta_\text{eqw}(L) \approx \beta_0 - \ln q/V \Delta \hat{u}$, where quantities provided with a caret are meant to be evaluated at the transition point. Analogously, one finds that the energetic Binder parameter $B=1-  \langle e^4 \rangle/3 \langle e^2 \rangle^2 $ has its local minimum $B_\text{min} \approx 1 - (\hat{u}_\text{o}/\hat{u}_\text{d}+\hat{u}_\text{d}/\hat{u}_\text{o})^2/12$ for $W_\text{o} \approx W_\text{d} \hat{u}_\text{d}^2 / \hat{u}_\text{o}^2$ at the inverse temperature $\beta_{B_\text{min}}(L) \approx \beta_0 - \ln \left( q \hat{u}_\text{o}^2/ \hat{u}_\text{d}^2 \right)/V \Delta \hat{u}$. 

Although this is a rather simple ansatz, it is able to reproduce the prefactors of the leading finite-size scaling corrections from a mathematically rigorous theory~\cite{Borgs1990,Borgs1991,Borgs1992} based on the work of Pirogov and Sinai,~\cite{Pirogov1975, Pirogov1976} which has later also been extended in order to apply to models like \eqref{eq:Hamiltonian}.~\cite{Bricmont1989} Neglecting corrections which are bounded by an exponentially decreasing function, this rigorous theory yields the following asymptotic expansions~\cite{Lee1991,Janke1993,Mueller2014,Janke2003} as $L~\to~\infty$:
\begin{eqnarray}
 \hspace{-0.4cm}  \beta_{c_\text{max}} (L)  & =  &    \beta_0 - \frac{\ln q}{V \Delta \hat{u}}  + \frac{ 
\frac{\Delta \hat{c}}{2\beta_0 \Delta \hat{u}} \left[ (\ln q)^2-12 \right]+4 }{\beta_0 V^2 \Delta \hat{u}^2}  \nonumber \\
 & & + \mathcal{O} \left(  \frac{(\ln q)^3}{V^3}  \right) , \label{eq:Pirogov_Sinai_TCmax} \\
 \hspace{-0.4cm}     \beta_\text{eqw} (L) &   = &  \beta_0 - \frac{\ln q}{V \Delta \hat{u}} + \frac{\Delta \hat{c} \,(\ln q)^2}{2 \beta_0^2 V^2 \Delta \hat{u}^3} + \mathcal{O}\left( \frac{(\ln q)^3}{V^3}  \right), \label{eq:Pirogov_Sinai_Teqw} \\
   \hspace{-0.4cm}   \beta_{B_\text{min}} (L) &   =  &   \beta_0 - \frac{\ln \left( q \hat{u}_\text{o}^2/\hat{u}_\text{d}^2 \right)}{V \Delta \hat{u}} + \frac{a}{V^2} + \mathcal{O} \left(  \frac{(\ln  q)^3  }{V^3}  \right), \label{eq:Pirogov_Sinai_TBmin} 
\end{eqnarray}
where $\Delta \hat{c} = \hat{c}_\text{d} - \hat{c}_\text{o}$, $ c_\text{o}= \frac{\mathrm{d}u_\text{o}}{\mathrm{d}T}$, $c_\text{d}= \frac{\mathrm{d}u_\text{d}}{\mathrm{d}T} $ and $a$ is an expression which can be written as $a=a_1 + a_2 \ln q + a_3  (\ln q)^2$ with constants $a_1$, $a_2$, $a_3$.~\cite{Janke1993,Mueller2014} From a double-Gaussian approximation~\cite{Binder1984, Challa1986, Peczak1989} of the canonical energy probability distribution, one can also derive the finite-size scaling formula~\cite{Mueller2014}$^,$\footnote{Note that the finite-size scaling formula for $\beta_\text{eqh}(L)$ in Ref.~\onlinecite{Mueller2014} contains typing errors which are corrected in the present paper.}
\begin{equation}
 \beta_\text{eqh}(L) = \beta_0 - \frac{ \ln \left( q \sqrt{\hat{c}_\text{d}/ \hat{c}_\text{o}} \right)}{V \Delta \hat{u}} + \mathcal{O} \left( \frac{(\ln q)^2}{V^2} \right)
 \label{eq:double_gaussian_ansatz_beta_eqh}
\end{equation}
for the inverse pseudo phase transition temperature where the two peaks of the canonical energy probability distribution have the same height. In the above formulas \eqref{eq:Pirogov_Sinai_TCmax}, \eqref{eq:Pirogov_Sinai_Teqw}, \eqref{eq:Pirogov_Sinai_TBmin}, and \eqref{eq:double_gaussian_ansatz_beta_eqh}, the degeneracy parameter $q$ is meant to be evaluated at the respective inverse pseudo phase transition temperature and the finite lattice size.

Assuming $q$ to be constant yields the standard finite-size scaling $\sim L^{-3}$ for all of the here regarded inverse pseudo phase transition temperatures. However, if the number of ordered phases $q$ which enters the above finite-size scaling formulas is exponential in $L$, that is 
\begin{equation}
q\left(\beta_0(L),L\right) = A e^{BL} 
\label{eq:q_non_standard_ansatz}
\end{equation}
with constants $A,B>0$,~\footnote{It might be also conceivable that the parameters $A$ and $B$ could even (weakly) depend on the lattice size $L$ which would lead to further modifications of the finite-size scaling laws. However, this will not be considered here since we are not aware of any theoretical predictions.} then the scaling is significantly altered to $ \sim L^{-2}$. This has already been observed for the plaquette-only gonihedric Ising model where $q=2^{3L}$ is known in the whole low-temperature phase.~\cite{Mueller2014} In the case of the fcc Ising antiferromagnet \eqref{eq:Hamiltonian} where $V=4L^3$, the equations \eqref{eq:Pirogov_Sinai_TCmax}, \eqref{eq:Pirogov_Sinai_Teqw}, \eqref{eq:Pirogov_Sinai_TBmin}, \eqref{eq:double_gaussian_ansatz_beta_eqh} would then read
\begin{eqnarray}
\label{eq:beta_Cmax_nonstandard_ansatz}
 \hspace{-0.4cm} \beta_{c_\text{max}}(L) & \approx & \beta_\text{eqw}(L)  =  \beta_0 - \frac{B}{4 \Delta \hat{u} L^2} - \frac{\ln A}{4 \Delta \hat{u} L^3}  \nonumber \\ 
 && + \frac{\Delta \hat{c} B^2}{32 \beta_0^2 \Delta \hat{u}^3 L^4} + \frac{\Delta \hat{c} B \ln A}{16 \beta_0^2 \Delta \hat{u}^3 L^5} + \mathcal{O} \left( \frac{1}{L^6} \right), \\
\label{eq:beta_Bmin_nonstandard_ansatz}
 \hspace{-0.4cm} \beta_{B_\text{min}}(L) & = & \beta_0 - \frac{B}{4 \Delta \hat{u}L^2} - \frac{\ln \left( A \hat{u}_\text{o}^2 / \hat{u}_\text{d}^2 \right)}{4 \Delta \hat{u} L^3} + \frac{a_3 B^2}{16 L^4} \nonumber \\
 & & + \frac{2 a_3 B \ln A + a_2 B}{16 L^5} + \mathcal{O} \left( \frac{1}{L^6} \right), \\
\label{eq:beta_eqh_nonstandard_ansatz}
 \hspace{-0.4cm} \beta_\text{eqh}(L) & = & \beta_0 - \frac{B}{4 \Delta \hat{u} L^2} -\frac{\ln \left( A \sqrt{\hat{c}_\text{d}/ \hat{c}_\text{o}} \right)}{4 \Delta \hat{u} L^3} \nonumber \\
 & &+ \mathcal{O} \left( \frac{1}{L^4} \right). 
\end{eqnarray}
The formulas for $\beta_{c_\text{max}}(L)$ and $\beta_\text{eqw}(L)$ coincide here up to the given order.

Unfortunately, the number $q$ of ordered phases contained in the finite-size scaling laws \eqref{eq:Pirogov_Sinai_TCmax}, \eqref{eq:Pirogov_Sinai_Teqw}, \eqref{eq:Pirogov_Sinai_TBmin}, \eqref{eq:double_gaussian_ansatz_beta_eqh} is unknown for the fcc Ising antiferromagnet as discussed in section~\ref{sec:Model}. Following the considerations of Mueller et al.,~\cite{Mueller2014,Mueller2014a} it is conceivable that for small enough $L$, the non-standard scaling $\sim L^{-2}$ given by \eqref{eq:beta_Cmax_nonstandard_ansatz}, \eqref{eq:beta_Bmin_nonstandard_ansatz}, \eqref{eq:beta_eqh_nonstandard_ansatz} applies whereas for large enough $L$, the standard scaling $\sim L^{-3}$ holds.

Similarly, for the extremal values $c_\text{max}$, $B_\text{min}$ of the specific heat and the energetic Binder parameter, one arrives at the formulas~\cite{Lee1991,Janke1993}
\begin{eqnarray}
\frac{c_\text{max}}{V} & = &  \left(  \frac{\beta_0 \Delta \hat{u}}{2}  \right)^2 + \frac{\left(  \Delta \hat{c} - \beta_0 \Delta \hat{u} \right) \ln q + \hat{c}_\text{d} + \hat{c}_\text{o}}{2V} \nonumber \\
&& + \mathcal{O} \left( \frac{(\ln q)^2}{V^2} \right) , \\
 B_\text{min} & = & 1 - \frac{1}{12} \left(  \frac{\hat{u}_\text{o}}{\hat{u}_\text{d}}  + \frac{\hat{u}_\text{d}}{\hat{u}_\text{o}}   \right)^2   + \frac{b}{V}  + \mathcal{O} \left( \frac{(\ln q)^2}{V^2} \right) , \quad
\end{eqnarray}
where $b=b_1 + b_2 \ln q$ with constants $b_1$, $b_2$ and $q$ is meant to be evaluated at $\beta_{c_\text{max}}(L)$ or $\beta_{B_\text{min}}(L)$, respectively, and the corresponding finite lattice size $L$. Consequently, for $c_\text{max}/V$ and $B_\text{min}$, the standard ansatz corresponding to a constant number $q$ leads to the scaling $ \sim L^{-3}$, whereas the non-standard ansatz~\eqref{eq:q_non_standard_ansatz} yields
\begin{eqnarray}
\frac{c_\text{max}}{V} & = & \left(  \frac{\beta_0 \Delta \hat{u}}{2}  \right)^2 + \frac{\left(  \Delta \hat{c} - \beta_0 \Delta \hat{u} \right) B}{8 L^2} \nonumber \\
&& + \frac{\left( \Delta \hat{c} - \beta_0 \Delta \hat{u} \right)  \ln A + \hat{c}_\text{d} + \hat{c}_\text{o} }{8 L^3} + \mathcal{O} \left(  \frac{1}{L^4} \right) , \quad \label{eq:c_max_nonstandard_ansatz} \\
B_\text{min} & = &  1 - \frac{1}{12} \left(  \frac{\hat{u}_\text{o}}{\hat{u}_\text{d}}  + \frac{\hat{u}_\text{d}}{\hat{u}_\text{o}}   \right)^2  + \frac{b_2 B}{4 L^2} + \frac{b_1 + b_2 \ln A}{4 L^3} \nonumber \\
&& + \mathcal{O} \left(  \frac{1}{L^4} \right) \label{eq:B_min_nonstandard_ansatz}
\end{eqnarray}
for $V=4L^3$, resulting in the transmuted scaling $\sim L^{-2}$.

\subsubsection{Multicanonical simulation}
A multicanonical ensemble~\cite{Berg1991,Berg1992,Janke1992a,Janke1998,Berg2000,Berg2002,Berg2003,Janke2003b,Berg2004} is an artificial statistical ensemble with the goal to optimize the performance of a simulation which is confronted with rare events. A possible application are first-order phase transitions~\footnote{Here only temperature-driven first-order phase transitions are discussed.} where canonical simulations show large autocorrelation times because of the suppressed region between the two peaks of the canonical energy probability distribution $P_\text{can}(E)$. A multicanonical simulation can overcome this problem by enhancing the sampling probability of the rare states corresponding to the valley of $P_\text{can}(E)$. For example, this can be achieved by a flat multicanonical energy probability distribution $P_\text{muca}(E)$ which implies the multicanonical probability distribution $p^\text{eq}_\text{muca}(\mu) $ of the microstates $\mu$ to be
\begin{equation}
p^\text{eq}_\text{muca}(\mu) = p^\text{eq}_\text{muca}(E(\mu)) \propto \frac{1}{\Omega(E)}
\label{eq:ideal_weights_for_flat_muca_histogram}
\end{equation}
where $\Omega(E)=\sum_{\mu} \delta_{E(\mu),E}$ denotes the density of states. Since the density of states is generally unknown, the desired multicanonical weights $w(E) \propto p^\text{eq}_\text{muca}(E)$ obeying \eqref{eq:ideal_weights_for_flat_muca_histogram} are approximated via a recursion (iteration). A possible stable recursion algorithm is described in Refs.~\onlinecite{Berg1996,Berg2003,Janke2003b,Janke2008}. As soon as the recursion yields a sufficiently good estimator of the desired multicanonical weights, the iteration process is terminated and the obtained weights are employed to perform a production run. From this production run, one can obtain estimators of the corresponding multicanonical expectation values $\langle \mathcal{O} \rangle_\text{muca}= \sum_\mu \mathcal{O}(\mu) p^\text{eq}_\text{muca}(\mu)$ of observables $\mathcal{O}$ in the usual way. The quantities of a canonical ensemble at inverse temperature $\beta$ can then be calculated via
\begin{eqnarray}
P(E) & \equiv & P_\text{can}(E)  \propto  P_\text{muca}(E) \frac{e^{-\beta E}}{w(E)}, \label{eq:reweighting_can_energy_probability_distribution} \\
\langle \mathcal{O} \rangle & \equiv & \langle \mathcal{O} \rangle_\text{can} =  \frac{ \left\langle \mathcal{O} \frac{e^{-\beta E} }{w(E)}  \right\rangle_\text{muca} }{\left\langle \frac{e^{-\beta E}}{w(E)} \right\rangle_\text{muca} }. \label{eq:reweighting_can_expectation_values}
\end{eqnarray}

As suggested by Zierenberg et al.,~\cite{Zierenberg2013} the multicanonical simulation can easily be parallelized in order to distribute the computational effort on several processing units.

\section{Numerical results} \label{sec:Numerical_results}
In the following, the units are chosen such that $k_B=1$ and $|J|=1$. 

\subsection{Simulation}
We performed a parallel multicanonical Monte Carlo simulation with $128$ processes adopting the Metropolis update algorithm.~\cite{Metropolis1953} In order to reduce the runtime, a flat multicanonical energy probability distribution was demanded only in a subset $I \subset [E_\text{min},E_\text{max}]=[-2V,6V]$ of the energy range~\footnote{The ground-state energy $E_\text{min}$ is derived in Ref.~\onlinecite{Danielian1961}.} while sampling energies outside of $I$ was avoided. Depending on the choice of $I$, this restricts the reweighting range, that is the range of inverse temperatures $\beta$ for which reliable results for the canonical ensemble can be obtained via \eqref{eq:reweighting_can_energy_probability_distribution}, \eqref{eq:reweighting_can_expectation_values}. We chose $I \approx [-1.99V,-V]$ which is sufficient for the analysis of the phase transition. A further speedup of the simulation was achieved by performing a linear extrapolation of the logarithmic multicanonical weights after each cycle of the iteration run. This extrapolation corresponds to an optimization of the initial multicanonical weights which were in the beginning chosen to be constant in $I$. The length of the single iteration cycles increased in the course of the iteration run so that first, a rough estimate of the desired multicanonical weights was obtained which was then fine-tuned. Finally, the iteration was terminated when the following three conditions were satisfied simultaneously. First, the estimated multicanonical energy probability distribution has to be sufficiently flat in $I$. That means that the relative deviations of its minimal and maximal value  from the average are both smaller than 30\%. Second, the statistical weight~\footnote{The statistical weight of an iteration cycle is a quantity used in the recursion algorithm described in Refs.~\onlinecite{Berg1996,Berg2003,Janke2003b,Janke2008}} of the last iteration cycle is below 20\% within $I$. Third, the Metropolis acceptance probabilities change by maximally 5\% compared to the previous iteration cycle.

In the production run, we performed for every lattice size $L$ at least $10^5$ sweeps per process but did not end the simulation until every process rendered at least 2 tunnel events itself and at least 10 tunnel events on average. A tunnel event is here defined to be realized when starting from an energy below or equal to the minimum of $I$, the simulation reaches an energy above or equal to the maximum of $I$, or vice versa. The initial spin configuration was set to be an AB structure ground state and the first 300 sweeps were excluded for the equilibration.

\begin{figure}
\includegraphics{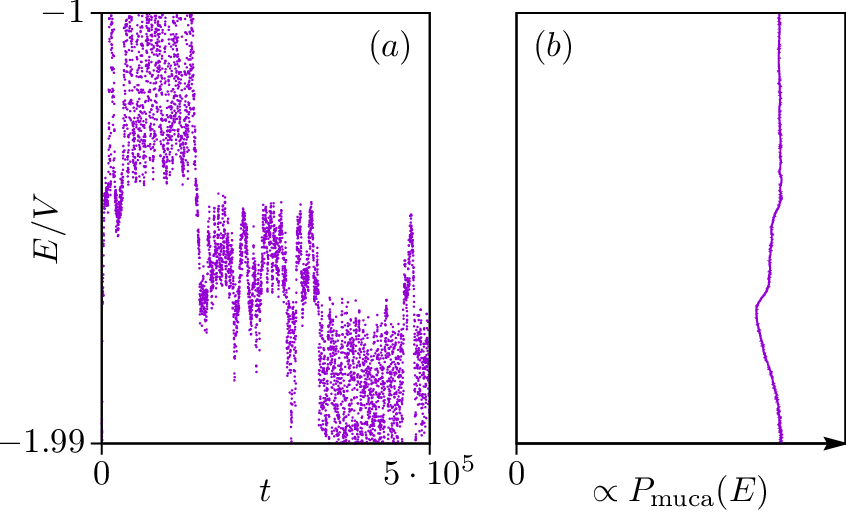}
 \caption{\label{fig:histogram_and_timeseries}(a)~Part of the time series of the energy per lattice site $e=E/V$ of one process of the production run for $L=16$. The variable $t$ denotes the sweep number. (b)~Corresponding estimated multicanonical energy probability distribution from all processes together. The vertical axis is the same as in the time-series plot.}
\end{figure}

We observed that despite of acceptably flat multicanonical energy probability distributions, the time series still show structures, see Fig.~\ref{fig:histogram_and_timeseries}. These structures probably correspond to remaining free-energy barriers in other ``directions'' than the energy, similar to the additional free-energy barriers originating from the droplet-formation~\cite{Nussbaumer2008,Nussbaumer2010} and the droplet-strip~\cite{Nussbaumer2010a} transitions in the two-dimensional Ising model.

\subsection{Analysis}
\begin{figure}
\includegraphics{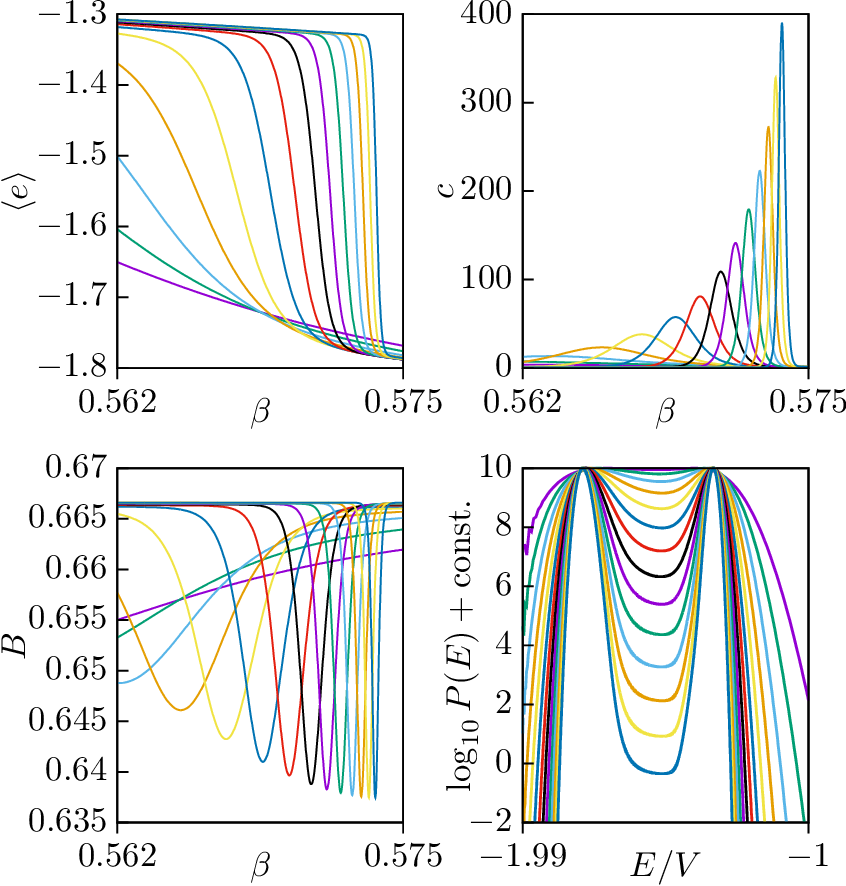}
 \caption{\label{fig:reweighting_curves}Estimated canonical quantities for the lattice sizes $L=5,\ldots,18$. The distributions $P(E)$ are here calculated at the respective inverse temperatures $\beta_\text{eqh}(L)$. For the sake of clarity, the error bars are not depicted here.}
\end{figure}

For the analysis of the simulated data, the Jackknife approach~\cite{Quenouille1956,Tukey1958,Miller1974,Efron1982} was applied to binning~\cite{Flyvbjerg1989} blocks which are represented by the different processes. From \eqref{eq:reweighting_can_energy_probability_distribution} and \eqref{eq:reweighting_can_expectation_values}, the canonical expectation value $\langle e \rangle$ of the energy per lattice site, the specific heat $c=\beta^2 V \left( \langle e^2 \rangle - \langle e \rangle^2 \right)$, the energetic Binder parameter $B = 1 - \langle e^4 \rangle/3 \langle e^2 \rangle^2$ and the canonical energy probability distribution $P(E)$ were estimated. The results are plotted in Fig.~\ref{fig:reweighting_curves}. One observes the typical behavior of a first-order phase transition where $\langle e \rangle(\beta)$ exhibits a discontinuity in the approach of the thermodynamic limit and $P(E)$ shows a double-peak structure near the phase transition point.

From these canonical quantities, the inverse pseudo phase transition temperatures $\beta_{c_\text{max}}(L)$, $\beta_{B_\text{min}}(L)$, $\beta_\text{eqw}(L)$, and $\beta_\text{eqh}(L)$ were determined. Here, $\beta_\text{eqw}(L)$ and $\beta_\text{eqh}(L)$ were calculated by the minimization of the functions $D_\text{eqw}(\beta) = |W_\text{o} - W_\text{d} |$ and $D_\text{eqh}(\beta) = | h_\text{o} - h_\text{d} |$, respectively, where the weights $W_\text{o/d}$ and the heights $h_\text{o/d}$ of the ordered and disordered peak of the canonical energy probability distribution $P(E)$ were estimated via $W_\text{o} = \sum_{E<E_\text{valley}} P(E) $, $W_\text{d} = \sum_{E \geq E_\text{valley}} P(E)$, $h_\text{o} = \max_{E< E_\text{valley}} P(E)$, $h_\text{d} = \max_{E \geq E_\text{valley}} P(E)$ with $E_\text{valley}$ denoting the location of the local minimum of $P(E)$ between the two peaks.

Via least squares fits,~\cite{Press2002,Young2015} the regarded inverse pseudo phase transition temperatures were fitted with the fit functions
\begin{eqnarray}
f^{(o)}_\text{n}(L) & = & k_0+\frac{k_1}{L^2}+\frac{k_2}{L^3}+\frac{k_3}{L^4}, \label{eq:non_standard_ansatz_fit_function} \\
f^{(o)}_\text{s}(L) & = & k_0+\frac{k_1}{L^3}+\frac{k_2}{L^6}+\frac{k_3}{L^9} , \label{eq:standard_ansatz_fit_function}
\end{eqnarray}
where the fit parameters $k_\alpha$ are set to zero for all $\alpha>o$ and $o=1,2,3$ denotes the order of the fit function. The functions $f^{(o)}_\text{n}$ correspond to the non-standard scaling according to \eqref{eq:beta_Cmax_nonstandard_ansatz}, \eqref{eq:beta_Bmin_nonstandard_ansatz}, \eqref{eq:beta_eqh_nonstandard_ansatz} whereas the functions $f^{(o)}_\text{s}$ represent the standard ansatz. For comparison, we fitted the data of the inverse pseudo phase transition temperatures also with the fit function
\begin{equation}
f_\text{BR}^{(1)}(L) = \left( k_0 + \frac{k_1}{L} \right)^{-1}
\end{equation}
which Beath and Ryan~\cite{Beath2006} employed, however, without a theoretical justification. 

\begin{figure}
\includegraphics{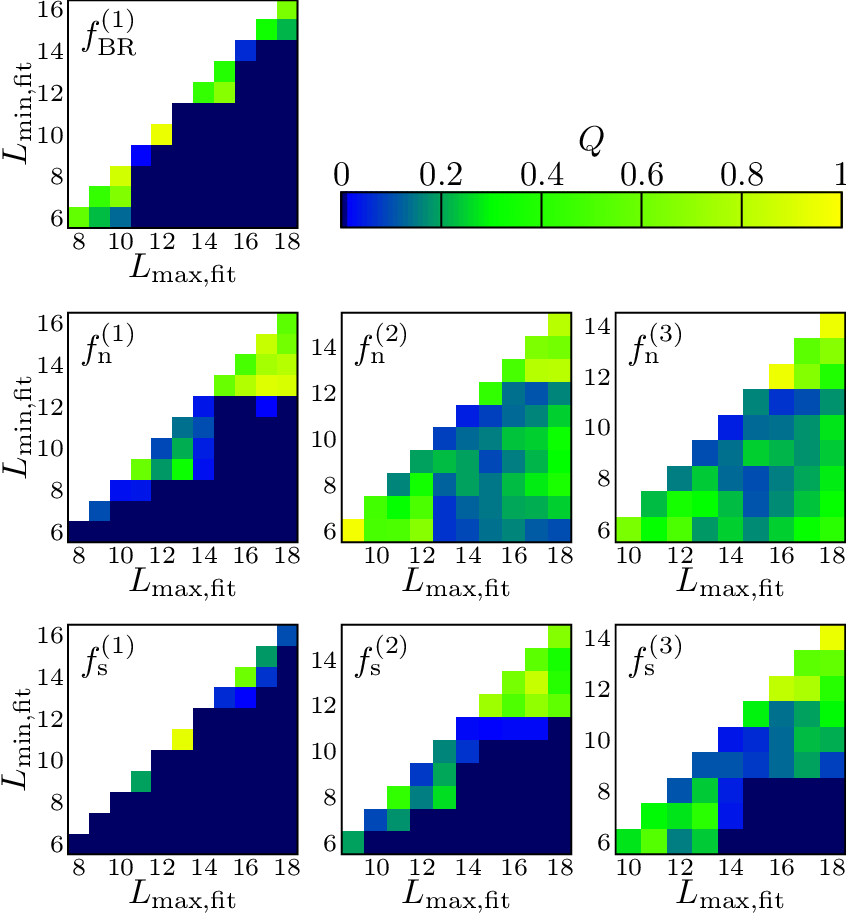}
 \caption{\label{fig:heatmaps}Quality of fit parameter $Q \in [0,1]$ for the fits of $\beta_\text{eqw}(L)$ with the different fit functions. Here, the lattice size range $[L_\text{min,fit},L_\text{max,fit}]$, for which the fit is performed, is varied. The color-value-identification of $Q$ is shown above.}
\end{figure}

Since the corrections arising from both, the truncation of the asymptotic expansions as well as the neglect of the additional exponentially bounded corrections~\cite{Janke1997} predicted by the Pirogov-Sinai theory are in general largest for the smallest lattice sizes, we varied the smallest lattice size $L_\text{min,fit}$ which is included in the fitting. In the same way, also the largest system size $L_\text{max,fit}$ included in the fit was varied in order to examine whether there is a crossover between the different scaling ansatzes. The results for the fits of $\beta_\text{eqw}(L)$ are shown in Fig.~\ref{fig:heatmaps} by means of the quality of fit parameter $Q$. The corresponding heatmaps for $\beta_{c_\text{max}}(L)$, $\beta_{B_\text{min}}(L)$, and $\beta_\text{eqh}(L)$ look very similar and hence are not depicted here.

One can see that up to the first order, the ansatz $f_\text{BR}^{(1)}$ of Beath and Ryan~\cite{Beath2006} is slightly better than the standard ansatz $f_\text{s}^{(1)}$ but clearly not as good as the non-standard ansatz $f_\text{n}^{(1)}$. Therefore, from now on, we will not examine the unjustified scaling ansatz $f_\text{BR}^{(1)}$ anymore. Taking into account also the second order, the non-standard ansatz $f_\text{n}^{(2)}$ yields good fits for all ranges $[L_\text{min,fit},L_\text{max,fit}]$ of fitted lattice sizes $L$. Of course, also the results of the standard ansatz $f_\text{s}^{(o)}$ improve with increasing order $o$, but they stay behind the non-standard ansatz $f_\text{n}^{(o)}$ in any case. This leads to the conclusion that up to the largest simulated lattice size $L=18$, the non-standard ansatz fits best. A crossover between the different ansatzes is not observable in the range of the simulated lattice sizes. However, it cannot be excluded that such a crossover might occur for larger lattice sizes $L>18$.

\begin{figure}
\includegraphics{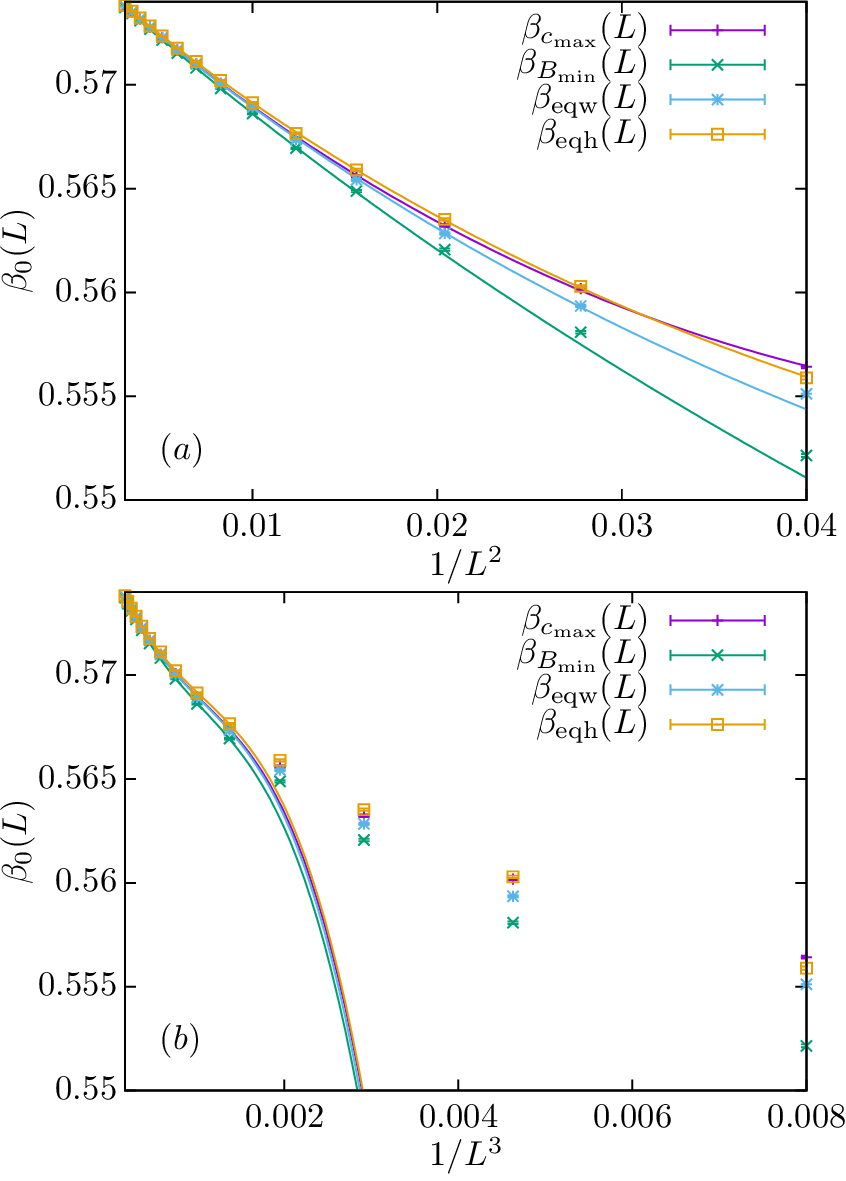}
 \caption{\label{fig:best_fit_betas}Obtained data of the regarded inverse pseudo phase transition temperatures for the lattice sizes $L=5,\ldots,18$ together with their best fits according to the non-standard ansatz~(a) and the standard ansatz~(b).}
\end{figure}

For each ansatz, the best fit is selected for $L_\text{max,fit}=18$. For the standard ansatz, the fit function up to order $o=3$ is chosen. For the non-standard ansatz and $\beta_{B_\text{min}}(L)$, $\beta_\text{eqw}(L)$, $\beta_\text{eqh}(L)$, it is order $o=2$ since $o=3$ does not significantly improve the quality of the fits, but results in comparably large relative statistical errors on the fit parameters. Solely for $\beta_{c_\text{max}}(L)$, order $o=3$ fits better in the non-standard ansatz. The best fit is then defined by the smallest value of $L_\text{min,fit}$ satisfying $Q>0.01$ and $\chi^2/N_\text{dof}<3$ where $\chi^2$ is the minimized quantity of the least squares fit and $N_\text{dof}$ is the corresponding number of degrees of freedom. In Fig.~\ref{fig:best_fit_betas}, the best fits of all regarded inverse pseudo phase transition temperatures for the two different ansatzes are shown. 

For both scaling ansatzes, each inverse pseudo phase transition temperature $\beta_{c_\text{max}}(L)$, $\beta_{B_\text{min}}(L)$, $\beta_\text{eqw}(L)$, $\beta_\text{eqh}(L)$ yields an estimate $\beta_0^{c_\text{max}}$, $\beta_0^{B_\text{min}}$, $\beta_0^\text{eqw}$, $\beta_0^\text{eqh}$ of the inverse phase transition temperature via the respective best fit. A final result $\beta_0$ is then obtained as the error weighted mean.~\footnote{To be more accurate, one would have to take into account also the cross-correlations between the different measurements like described in Refs.~\onlinecite{Weigel2009,Weigel2010}.}$^,$\cite{Weigel2009,Weigel2010} Its statistical error is calculated as the maximum of the corresponding error weighted standard deviation and the maximal error due to error propagation. The obtained values for both ansatzes are given in Table~\ref{tab:results}. According to their error bars, the final estimates of the phase transition point for the non-standard and the standard scaling ansatz differ significantly from each other which justifies their differentiation. As reasoned above, the non-standard ansatz should be employed. Its final estimate $T_0=1.735047(46)$ of the phase transition temperature is significantly below the previously commonly accepted value $T_0=1.76$ of Binder~\cite{Binder1980} but above the high-precision estimate $T_0=1.7217(8)$ of Beath and Ryan~\cite{Beath2006} which is not surprising since they used a different scaling ansatz.

\begin{table}
 \caption{\label{tab:results}Resulting estimates of the phase transition point for the non-standard and the standard scaling ansatz. The estimate $\beta_0$ is calculated by a combination of $\beta_0^{c_\text{max}}$, $\beta_0^{B_\text{min}}$, $\beta_0^{\text{eqw}}$, and $\beta_0^{\text{eqh}}$ which in turn are obtained from the best fits of the respective inverse pseudo phase transition temperatures. Besides, $T_0=1/\beta_0$ is computed where the corresponding error is determined via maximal error propagation.}
  \begin{ruledtabular}
\begin{tabular}{l | d{1.9} d{1.9}}
 & \multicolumn{1}{c}{Non-standard ansatz} & \multicolumn{1}{c}{Standard ansatz} \\ \hline
 $\beta_0^{c_\text{max}}$ & 0.576338(31) & 0.575432(24) \\
 $\beta_0^{B_\text{min}}$ & 0.576363(14) & 0.575432(24) \\
 $\beta_0^\text{eqw}$ & 0.576358(16) & 0.575431(23) \\
 $\beta_0^\text{eqh}$ & 0.576342(14) & 0.575436(24) \\
 $\beta_0$ & \multicolumn{1}{Z{.}{.}{1.9} }{0.576353(15)} & \multicolumn{1}{Z{.}{.}{1.9} }{0.575433(24)} \\
 $T_0$ & \multicolumn{1}{Z{.}{.}{1.9} }{1.735047(46)} & \multicolumn{1}{Z{.}{.}{1.9} }{1.737823(70)}
\end{tabular}
 \end{ruledtabular}
\end{table}

\begin{figure}
\includegraphics{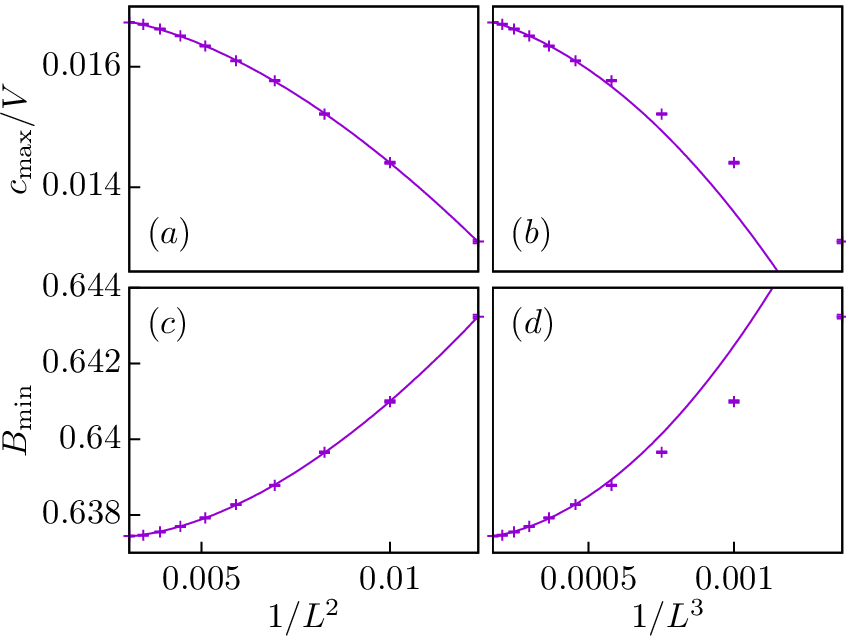}
 \caption{\label{fig:best_fit_Cmax_Bmin}Obtained data of $c_\text{max}/V$ (upper plots $(a)$, $(b)$) and $B_\text{min}$ (lower plots $(c)$, $(d)$) for the lattice sizes $L=9,\ldots,18$ together with their best fits according to the non-standard (plots $(a)$, $(c)$ on the left) and the standard ansatz (plots $(b)$, $(d)$ on the right).}
\end{figure}

We also determined $c_\text{max}/V$ and $B_\text{min}$, and fitted them up to order $o=2$ with the fit functions $f_\text{n}^{(o)}$ and $f_\text{s}^{(o)}$ which correspond to the non-standard ansatz \eqref{eq:c_max_nonstandard_ansatz}, \eqref{eq:B_min_nonstandard_ansatz} and the standard ansatz, respectively. As described for the inverse pseudo phase transition temperatures, the best fits are selected, which is here achieved for $o=2$. The results are shown in Fig.~\ref{fig:best_fit_Cmax_Bmin}. Also here, one can see that the non-standard ansatz fits better then the standard ansatz.

\vspace{-0.5cm}

\section{Conclusion} \label{sec:Conclusion}
We studied the phase transition of the fcc Ising antiferromagnet \eqref{eq:Hamiltonian} via a parallelized Monte Carlo simulation in a multicanonical ensemble. First of all, it could be confirmed that the transition is of first order since we observed a clear double-peak structure of the canonical energy probability distribution $P(E)$. Besides, we found that after bypassing the free-energy barrier which causes the valley between the two peaks of $P(E)$, there are still other ``hidden'' barriers left.

The main focus of the investigation lay on the finite-size scaling analysis of the common inverse pseudo phase transition temperatures $\beta_{c_\text{max}}(L)$, $\beta_{B_\text{min}}(L)$, $\beta_\text{eqw}(L)$, $\beta_\text{eqh}(L)$ and also of the extremal values $c_\text{max}/V$, $B_\text{min}$ of the specific heat per lattice site and the energetic Binder parameter. For the simulated lattice sizes $L \leq 18$, all of these quantities complied with the non-standard scaling $\sim L^{-2}$ instead of the standard scaling $\sim L^{-3}$ like it was previously also observed for the plaquette-only gonihedric Ising model. Employing this non-standard scaling ansatz, we obtained the value $T_0=1.735047(46)$ for the phase transition temperature.

The transmutation of the finite-size scaling is probably caused by the exponential ground-state degeneracy of the fcc Ising antiferromagnet. A crossover to the standard scaling for increasing $L$ was not observed. However, this could happen for larger systems $L > 18$ which would be an interesting, but computationally very demanding question for a future project.

\begin{acknowledgments}
We thank Marco Mueller for useful discussions. This work was in part supported by the Deutsch-Franz\"osische Hochschule (DFH-UFA) through the Doctoral College ``${\mathbb L}^4$'' under Grant No.\ CDFA-02-07 and the Leipzig Graduate School of Natural Sciences ``BuildMoNa''.
\end{acknowledgments}

\end{document}